\documentclass[11pt]{article}
\font\cero=cmss10 scaled 1728
\font\uno=cmssbx10 scaled 1200

\usepackage{tensor}
\usepackage[usenames]{color}
\usepackage{verbatim}
\usepackage{amsfonts}

\usepackage{graphicx}
\usepackage{float}
\begin{document}
\begin{flushleft}
{\cero The Higgs mechanism and geometrical flows for two-manifolds} \\
\end{flushleft}
{\sf R. Cartas-Fuentevilla, A. Herrera-Aguilar}\\
{\it Instituto de F\'{\i}sica, Universidad Aut\'onoma de Puebla,
Apartado postal J-48, Puebla, Pue., 72570, M\'exico}.\\
{\sf J. Berra-Montiel}\\
{\it Facultad de Ciencias, Universidad Aut\'onoma de San Luis Potos\'i,
Campus Pedregal, Av. Parque \\
Chapultepec 1610, Col. Privadas del Pedregal, San
Luis Potos\'i, SLP, 78217, M\'exico.}\\ 

Using Perelman's approach for geometrical flows in terms of an entropy functional, the Higgs mechanism is studied
dynamically along flows defined in the space of parameters and in fields space. The model corresponds to two-dimensional gravity that incorporates torsion as the gradient of a Higgs field, and with the reflection symmetry to be spontaneously broken. The results show a discrete mass spectrum, and the existence of a mass gap between the Unbroken Exact Symmetry and the Spontaneously Broken Symmetry scenarios. In the later scenario, the geometries at the degenerate vacua correspond to conformally flat manifolds without torsion; {\it twisted} two-dimensional geometries are obtained by building perturbation theory around a ground state; the tunneling quantum probability between vacua is determined along the flows.\\

\noindent KEYWORDS: geometrical flows, Higgs mechanism, torsion, two-manifolds.\\
PACS numbers: 11.10.Hi, 04.90.+e, 11.15.Ex, 04.50.Kd

\section{Introduction}

As a geometrical evolution equation, the Ricci flow \cite{Hamilton} has attracted the attention of the physical community since it behaves as a heat equation for the metric, homogenizing the metric on a given manifold. 
The most obvious applications of the Ricci flow in physics can be realized within general relativity, since it is a theory about the geometry of space and time. 

Nowadays, the Ricci flow has been applied to study several problems within the framework of physics: 
A possible relation between Perelman's entropy and the geometric Bekenstein-Hawking entropy from black hole thermodynamics was considered at \cite{GFBHE} based on a study of the fixed points of the flow. It appeared that Perelman's entropy has no connection to the geometric entropy, however, a modified flow does appear to relate both entropies.
Further, a remarkable relation between the evolution of the area of a closed surface and the corresponding Hawking mass under the Ricci flow was obtained in \cite{EnergyERF}, revealing that the rate of change of the area is bounded by its Hawking mass. 
The behavior of the ADM mass of an asymptotic locally Euclidean (ALE) space along the Ricci flow was investigated in \cite{Dai}, revealing that the mass is invariant under the flow in three dimensions and  that  the ALE property is conserved under the Ricci flow.
The Ricci flow was also applied to $4d$ Euclidean gravity with a $S^1\times S^2$ boundary, realizing the canonical ensemble for gravity in a box \cite{wiseman}; it turned out that at high temperature the action possesses three saddle points, one unstable under the Ricci flow, and other two that respectively lead to a large black hole and to a hot flat space (via a topology-changing singularity in this case). 
The Ricci flow has been implemented as well to deform wormhole geometries, yielding three scenarios for the evolution of the wormhole throat (shrinking, expanding or steady) determined by a critical parameter that also reveals topological changes on a manifold \cite{Viqar}.

A review on how the Ricci flow arises in the renormalization group (RG) of non-linear sigma models was presented in \cite{woolgar}, along with a relation between the Ricci flow and Ricci solitons on the basis of the behavior of the mass under the flow; moreover, a 
discussion about the construction of $4d$ Lorentzian and Euclidean analytic Ricci solitons from a $3d$ seed solution (with the aid of a scalar field) was given. 
Subsequently, the analysis of the asymptotically hyperbolic mass under the curvature-normalized Ricci flow of conformally compactifiable and asymptotically hyperbolic manifolds was performed in \cite{BahelowskyWoolgar}, showing that the mass exponentially vanishes along the flow for $d\geq 3$ manifolds.
By using a maximum principle, it was proved that Ricci solitons do not exist within static Lorentzian spacetimes which are asymptotically flat, Kaluza-Klein, locally AdS or have extremal horizons \cite{figueras}. 
Additionally, the connection between the holographic RG flow and the Ricci flow was revealed in \cite{kiritsis,jackson}. In particular, the Hamilton-Jacobi formalism was introduced in \cite{jackson} in order to derive the RG flow 
equations and to show that for dual AdS/QFT theories, the RG flow is described by the Ricci flow. 	
Besides, a proof of the irreversibility of a wide class of world-sheet RG flows to first order in $\alpha'$ in string theory for asymptotically flat target manifolds was reported in \cite{OSW1}-\cite{OSW2}. 
One more physical theory in which the Ricci flow emerges in a natural way is the Ho\v{r}ava-Lifshitz gravity \cite{horava,nishioka}, where the Hamilton Ricci flow arises as a limit of a generalized RG flow under the application of the Hamilton-Jacobi equation to the Ho\v{r}ava-Lifshitz theory of gravity. 

One more appearance of the Ricci flow as a RG flow within the field theory realm takes place when the flow parameter encodes the observation scale of the theory under consideration \cite{Chowdhury}. Thus, a change in scale corresponds to a shift in the magnification of a microscope used to study a given physical system: The bigger the magnification, the more detailed the obtained image is and the more information we get. One can physically interpret the decrease of information as an increase of entropy, in this sense the Ricci flow diminishes the information of a given manifold and leaves just the topological information needed to identify the manifold \cite{Chowdhury,perelman}. 

It is well-known that flat spaces are fixed points of the Ricci flow for some geometries \cite{Oliynyk}, for instance, the evolution of the Ricci flow under specific initial conditions (e. g. rotationally symmetric and asymptotically flat initial data) is immortal in the sense that the solution exists for the whole interval $(\lambda_0,\infty)$, remains asymptotically flat,  and asymptotically converges to flat Euclidean space as the flow parameter goes to infinity. There are more complicated geometries, such as the direct-product metrics $(\mathcal{T}^2, \mu) \times (S^1, dx^2)$ where $\mu$ is an arbitrary Riemannian metric on $\mathcal{T}^2$, that also converge to flat metrics \cite{RStability}.
More recently, a study of the behavior of maximally symmetric manifolds with constant curvature, de Sitter (dS) and Anti de Sitter (AdS) spaces, under the so-called Hamilton-DeTurck Ricci flow\footnote{
An equivalent formulation of the Ricci flow was introduced by D. DeTurck \cite{Turck} through a family of diffeomorphisms 
along the flow, giving rise to the Hamilton-DeTurck Ricci flow.} was reported in \cite{CFHAOS} by analytically solving the modified Ricci flow equations with a simple ansatz for the DeTurck vector field and revealing a change in the signature of the metric at a singular point in the curvature, a transition from positive to negative curvature throughout this singularity, and a further evolution towards Minkowski spacetime. One more family of exact solutions to the Hamilton-DeTurck Ricci flow was obtained for Lifshitz spacetimes represented by metrics that are invariant under anisotropic scalings of space and time \cite{CHAHM}; these 
spaces play a very important role in the description of holographic quantum systems within the framework of the Gravity (Lifshitz)/Condensed Matter Theory correspondence \cite{Taylor,HLS}. These exact solutions to the Ricci flow equations show as well that Lifshitz spaces with discrete and continuous critical exponents tend to a fixed point that corresponds to Minkowski spacetime as the flow parameter evolves towards infinity. Thus, the Ricci flow generally homogenizes the metric in such a way that it reaches the equilibrium state represented by flat spacetime.

\section{Motivation}

Here we would like to motivate our work from the connection of two intrinsically different perspectives on the Ricci flow: physics and mathematics. On the one hand, the recent discovery of the Standard Model Higgs boson reported by both the ATLAS and the CMS collaborations \cite{cms-atlas}, and on the other, the final proof of the Poincar\'e conjecture given by G. Perelman \cite{perelman}, milestones that represent a breakthrough in these research areas. The first confirms the existence of the only elementary particle in the Standard Model that has not yet been observed, the second unravels one of the most fundamental unsolved problems in mathematics for close to a century. The relevance of the first is clear, the Standard Model describes successfully high energy particle interactions, and the detection of the Higgs boson represented also a scientific goal of the Large Hadron Collider. On the other hand, the longstanding Poincar\'e conjecture is only a consequence of a more general result proved by Perelman on geometrization theorems, overcoming the difficulties in the programme of R. S. Hamilton, which uses the Ricci flows as the fundamental tool \cite{Hamilton}. Such theorems have a potential importance in physics, beyond the obvious impact in mathematics; for example, in the path-integral quantization for geometrical theories, one must sum over topologies and geometries, and a criterion for discerning over inequivalent histories is mandatory. The geometrical flows appear also in field theory by describing the RG equations, providing an approach to the problem of tachyon condensation and vacuum selection in string theory, where it has been shown that there exists an action functional whose gradient (with respect to a metric on the space of coupling constants) produces the RG flow of the target space metric in the worldsheet sigma model to all orders in perturbation theory \cite{Tseytlin}. This fact yields a monotonicity formula for the flow which fails only when the perturbation series of the beta function does not converge, a situation that arises when the curvatures or their derivatives grow large at next-to-leading order in perturbation theory \cite{OSW3}.

It is not evident that there exists a direct relationship between both scenarios, and the purpose of the present paper is to show that the fundamental role played by the Higgs mechanism in the Standard Model can be extended to the context of the geometrical flows and, conversely, to show the behavior of that mechanism of symmetry breaking under flows defined in the field space; the results will reveal qualitatively new aspects in both senses.
In order to simplify the discussion and ideas, we focus on two-dimensional manifolds describing gravity; the dynamical variables
are the metric and the torsion, which is identified with the gradient of a scalar Higgs field $\phi$. The respective action, identified as an entropy functional in field space within the context of the geometrical flows, corresponds to the scalar curvature of the manifold, which mimics the kinetic term for the Higgs field; such Lagrangian term will be supplemented with a Higgs potential; the full action has the discrete symmetry $\phi \rightarrow -\phi$, which will be spontaneously broken.

The paper is further organized as follows: In Section 3 the action, conventions, and the basic formalism, are briefly exposed. In Section 4, the metric, Higgs, and mass flows are constructed by the requirement of a non-decreasing entropy functional along the flows. The mass flow is solved explicitly, and we show in details the qualitative behavior. Section 5 is devoted to the description of vacuum in both the Unbroken Exact symmetry (UES) case, and most importantly, in the Spontaneous Symmetry Breaking (SSB) scenario, for which the evolution of the Higgs potential and the tunneling probability between vacua are studied.

\section{Preliminaries and entropy functional}

In this part, as a first step we shall consider that the geometry of a manifold is encoded in the metric compatible only with the symmetric part of the connection as regarded in \cite{cartas}, this approach allows us to introduce torsion and to retain some properties of the usual Riemannian manifolds with a Levi-Civita connection. If the full covariant derivative ($\nabla$) of the metric is decomposed according to the decomposition of the full connection in terms of symmetric and anti-symmetric parts in the lower indices $\Gamma^{i}_{jk}  = \widehat{\Gamma}^{i}_{(jk)} + \frac{1}{2}T^{i}_{jk}$, as $\nabla_{i} g_{jk}=\widehat{\nabla}_{i} g_{jk}-\frac{1}{2}T^{l}_{ji}g_{lk}-\frac{1}{2}T^{l}_{ki}g_{jl}$, then the compatibility condition of the metric with respect to $\widehat{\Gamma}$ alone is achieved if the tensorial constraint $\widehat{\nabla}_{i} g_{jk}=0$ is satisfied, which leads to
$\widehat{\Gamma}^{k}_{(ij)}=\frac{1}{2}g^{kl}[\partial_{i} g_{jl}+\partial_{j} g_{il}-\partial_{l} g_{ij}]$. Note that this approach can be interpreted as a variant of the Ricci theorem, and does not involve the {\it contorsion} (for more details see \cite{cartas}); moreover, within the present framework, the torsion will be determined dynamically though an identification with the Higgs field.
The full curvature of the two-manifold can be displayed as
\begin{equation}
R^{i}{_{jkl}}(\Gamma =\widehat{\Gamma}+\frac{1}{2}T) =\widehat{R}^{i}{_{jkl}}(\widehat{\Gamma}) +
S^{i}{_{jkl}}(\widehat{\Gamma},T), \label{cur-total} \end{equation}
where $\widehat{R}^{i}{_{jkl}}(\widehat{\Gamma}(g))=
\delta^{i}{_{[k}}g_{l]j}\widehat{R}$, for the particular case of two dimensions, and
\begin{equation}
     S^{i}{_{jkl}} = \delta^{i}_{j}F_{kl} + \delta^{i}_{[k}F_{l]j} + 2\delta^{i}_{[k}S^{+}_{l]j} - 2\delta^{i}_{[k}g_{l]j}S,
     \label{S-curvature}
\end{equation}
with
\begin{eqnarray}
     S^{+}_{ij} \!\! & \equiv & \!\! g^{kl}S_{iklj} = -\frac{1}{2} (\widehat{\nabla}_{(i}T_{j)} - \frac{1}{2}T_{i}T_{j}) + \frac{1}{2} g_{ij} (\widehat{\nabla}_{k} - \frac{1}{2}T_{k}) T^{k}, \nonumber \\
     S \!\! & = & \!\! g^{ij}S^{+}_{ij} = \frac{1}{2} (\widehat{\nabla}_{k} - \frac{1}{2}T_{k}) T^{k}; \qquad     F_{ij}  = \partial_{[j} T_{i]},                                       \label{traces}
\end{eqnarray}
where we have considered that in two-dimensions the torsion is fully described by its trace $T_{i}\equiv \delta^{k}_{j} T_{i}{^{j}}{_{k}}$. Besides the compatibility of the metric with $\widehat{\Gamma}$, now we make the second assumption within our approach; in what follows we identify the torsion vector field with the gradient of a neutral scalar field such as $T_{i}=\partial_{i}\phi$, leading trivially to $F_{ij} = 0$, and to the following relations
\begin{eqnarray}
S^{i}{_{jkl}} \!\! & = & \!\! 2 \delta^{i}{_{[k}} R^{+}_{l]j}, \quad
     S^{+}_{ij}  =  \frac{1}{2} g_{ij} (\widehat{\nabla}_{k} - \frac{1}{2}\partial_{k}\phi ) \partial^{k}\phi + R^{+}_{ij}, \nonumber \\
          R^{+}_{ij} \!\! & = & \!\! -\frac{1}{2} \widehat{\nabla}_{(i} \widehat{\nabla}_{j)}\phi + \frac{1}{4} \partial_{i}\phi\cdot\partial_{j}\phi, \quad
          S = - \frac{1}{4} \partial_{i}\phi\cdot\partial^{i}\phi + \frac{1}{2} \widehat{\nabla}_{i}\partial^{i}\phi.               \label{Higgs-S}
\end{eqnarray}
The full scalar curvature of the two-manifold $\widehat{R}-S$, proves to be a geometrical invariant from which we shall construct an entropy functional; once we have identified the torsion with the gradient of a Higgs field, we can add a Higgs potential with a $``\lambda \phi^{4}"$ interaction, leading to the functional
\begin{equation}
     {\cal E} (g,T) \equiv \int_{M} \sqrt{g} (S-\widehat{R}) d^{2}x-\int_{M} \sqrt{g} V d^{2}x,
     \label{full-action}
\end{equation}
where $V=-\frac{1}{2} m^{2}\phi^{2} - \frac{1}{16}\lambda\phi^{4}$, with $\lambda> 0$; and in general, the expressions associated to $m$ and $\lambda$ depend on the parameter $t$, as $m=m(t)$, and $\lambda = \lambda (t)$. Note that with the identification $T=\nabla\phi$, the $S$-curvature corresponding of the manifold $M$ reduces, modulo a total derivative, to the {\it kinetic} term of the Higgs-like field; thus, for two-manifolds without boundaries, we can observe that the entropy functional $\cal E$ has the reflection symmetry $\phi\rightarrow -\phi$. In appearance, this action looks like pure gravity ($\hat{R}$) coupled to a scalar field ($S$); however the geometrical content is different, since the complete scalar curvature $\hat{R}-S$ is ``pure gravity", {\it i.e.} vacuum two-dimensional gravity, and in this sense the geometry of the two-manifold is exotic. Additionally the term $\hat{R}$ is topological in two -dimensions, which means that the dynamics will be generated only by the term $S$ through the field $\phi$; variationally the term $S$ leads to the usual wave equations of motion for the scalar field $\phi$, but in this context they correspond geometrically to the Einstein (vacuum) equations of motion. Therefore, $S$ corresponds to a  {\it massless} Lagrangian term, and the inclusion of the potential can be considered as mass corrections.
\section{Metric, Higgs, and mass flows}
Perelman flows are strictly defined only in the Euclidean framework; thus in the following we work on a two-dimensional manifold $M$ with a positive definite metric; the ``time" parameter along the flow is denoted by $t$.
The variation of the entropy functional is given by
\begin{eqnarray}
     \partial_{t}{\cal E}\!\! & = & \!\! \frac{1}{4}\int_{M}\sqrt{g}\{ [\partial^{i}\phi\cdot\partial^{j}\phi+4\widehat{R}^{ij}- \frac{1}{2} g^{ij}((\partial\phi)^{2}+4\widehat{R})]-2Vg^{ij}\}\partial_{t}g_{ij}\nonumber
      \\
      \!\! & + & \!\!\int_{M}\sqrt{g} [(\frac{1}{2} \widehat{\Box}\phi - \frac{\partial V}{\partial\phi}) \partial_{t}\phi + \frac{1}{2}\phi^{2} \partial_{t}m^{2} + \frac{1}{16}\phi^{4} \partial_{t}\lambda], \label{var-2}
\end{eqnarray}
where $\widehat{\Box}\equiv\widehat{\nabla}_{i}\partial^{i}$ is the two-dimensional Laplace-Beltrami operator; and also the metric variations have been separated in a traceless part and the trace depending on the Higgs potential.
Now, we assume that the metric of the two-manifold (with torsion) is conformally flat, that is $g_{ij}\equiv\delta_{ij}\Omega^{2}(x,t)$; thus, the traceless part vanishes trivially, reducing the flow for the conformal factor to
\begin{equation}
     \partial_{t} \Omega^{2} = -\omega V \Omega^{2l}, \quad l = 0,\pm 1,\pm 3 ,\pm 5,...\label{metric-flow}
\end{equation}
with the constant $\omega\geq0$; under this flow the metric variations of the entropy reduce to $\omega V^2 \Omega^{2(l-1)}$, which results to be strictly positive. Note that the inclusion of the Higgs potential makes dynamical the conformal factor along geometrical flows; thus, the stationary points are characterized by $V=0$ ({\it i.e.}, $\phi=0$), or by the condition $\omega=0$, which corresponds to the trivial case.
The change of the area $\cal A$ associated to the manifold $M$ is controlled by,
\begin{equation}
     \partial_{t} {\cal A} = \frac{1}{2} \int_{M} d^{2}x \sqrt{g} g^{ij} \partial_{t} g_{ij} = -\omega\int_{M}\sqrt{g}V\Omega^{2(l-1)} d^{2}x = -\omega\int_{M}V\Omega^{2l} d^{2}x  ;
     \label{area}
\end{equation}
since $\omega\Omega^{2l}$ is strictly positive, we observe that the Higgs potential will determine a qualitatively different behavior of the area in the UES and SSB scenarios; in particular for $l=0$, the gradient of ${\cal A}$ is governed by the average of the Higgs potential over $M$.
Similarly the Higgs flow must be
\begin{equation}
     \partial_{t}\phi = \frac{1}{2} \widehat{\Box}\phi - \frac{\partial V}{\partial\phi};
     \label{Higgs-flow}
\end{equation}
which corresponds to a diffusion-like equation for the Higgs field; note that both fields $\phi$, and $\Omega$ appear coupled in the evolution Eqs. (\ref{metric-flow}), and (\ref{Higgs-flow}); in general the stationary points for the  Higgs flow satisfy $\frac{1}{2} \widehat{\Box}\phi - \frac{\partial V}{\partial\phi}=0$, and define also the space of solutions of the equations of motion.

Additionally, the monotonicity of ${\cal E}$ requires that $\partial_{t}m^{2}\geq 0$, and $\partial_{t}\lambda\geq 0$; the cases corresponding with strict equalities prove to be trivial, since they imply that $m^{2}$ and $\lambda$ are stationary along the flows. The UES and SSB scenarios require that $\lambda >0$; hence, if this restriction is imposed at $t=0$, then the condition $\partial_{t}\lambda \geq 0$ required by the monotonicity of ${\cal E}$ will maintain the positivity of $\lambda$ for all $t$, and the scenario for the Higgs symmetry-breaking is possible along the flows.
As one may expect, the more delicate aspect is the behavior  of the mass along the flows; the UES scenario requires $m^{2}>0$, and SSB requires $m^{2}<0$; however, the monotonicity will be consistent with both, since the restriction $\partial_{t}m^{2}\geq0$ can be realized through the mass flow
\begin{equation}
     \partial_{t}m^{2} = \mu (m^{2})^{n}, \qquad m^{2}(t) = [(n-1) (-\mu t+ \mu_{0})]^{\frac{1}{1-n}},\quad n=0, \pm 2, \pm 4, \ldots
     \label{mass-flow}
\end{equation}
where $\mu$ is a constant such that $\mu \geq 0$; the solution of this flow is displayed explicitly in Figure 1, where
 $\mu_{0}$, corresponds to an arbitrary constant but without positivity restrictions. We can choose $t=0$ as the moment at which the usual SSB occurs; eventually the system is left to evolve along the flow. There exist two scenarios for the mass spectrum at $t=0$, depending
on the sign of $\mu_{0}$. In the case $\mu_{0}<0$, we have that $m^{2}(0)> 0$ for $n \leq 0$, and $m^{2}(0)< 0$ for $n \geq 2$; similarly for $\mu_{0}>0$, we have that $m^{2}(0)>$ for $n \geq 2$, and  $m^{2}(0)<0$ for $n \leq 0$. However, the two scenarios are qualitatively equivalent, and for the sake of convenience we choose the first one for discussion; in the following Figure we show the profile of $m^{2}(0)$ against $n$, according to the Eq. (\ref{mass-flow}).
\begin{figure}[H]
  \begin{center}
    \includegraphics[width=.5\textwidth]{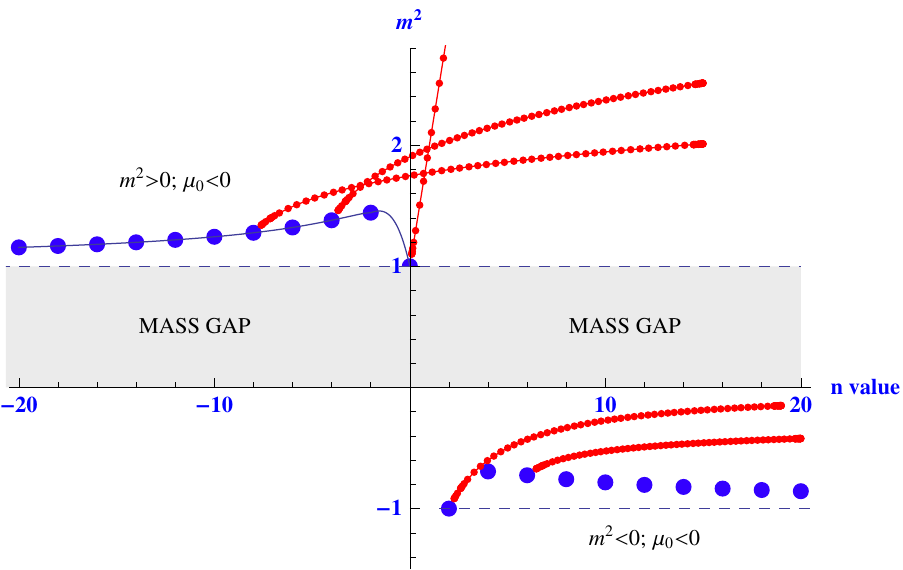}
  \caption{Each blue point represents the (real) value of $m^{2}(0)$ for $n$ in Eq. (\ref{mass-flow}); for $n \leq 0$ the mass spectrum is continuous, since all values of $m^{2}(0)$ are real, this fact is shown by a continuous curve connecting all these points;  this region in the second quadrant correspond to UES. For  $n \geq 2$ in the fourth quadrant, between two blue points corresponding to real values, there exist in general complex values for $m^{2}(0)$; thus, there is not a continuous curve connecting these points, then, it corresponds to the SSB with a strict discrete mass spectrum. In appropriate units, we can fix $m^{2}(t=0, n=0)=1$, and  $m^{2}(t=0, n=2)=-1$; additionally we can see that $lim_{n\rightarrow \pm \infty} m^{2}(t=0,n)=\mp 1$, which is indicated by blue horizontal dashed lines.}
  \label{fig:ejemplo}
  \end{center}
\end{figure}
Once we have described the profile of mass at $t=0$, we left it to evolve according to the  Eq. (\ref{mass-flow}), where in the figure we can understand the "x-axis" as the $t$-axis; then $m^{2}(t)$ evolves for each value of $n$, as indicated by the red (meshed) lines, in appropriate units of $\mu$.  These red lines go always to the right, while the mass is increasing, showing the restriction $\partial_{t}m^{2}>0$; thus the scenario of UES seems to be stable along the flow, with $lim_{t\rightarrow  +\infty} m^{2}(t,n)=+\infty$.  Similarly in the fourth quadrant we have that $lim_{t\rightarrow  +\infty} m^{2}(t,n)=0$,  and the red lines never touch the "x-axis" at finite time; thus, the SSB scenario seems also to be stable along the mass flow. Note that in fact there exists a Mass Gap region between the two scenarios, which is inaccessible for the mass flow; the gap is increasing along the mass flow. Therefore the UES and SSB scenarios seem to be disconnected in the parameters space.

The  case with $\mu_{0}>0$ can be obtained from the Figure 1 by reflecting the blue points though the "x-axis"; the effect is that the UES and SSB are interchanged; in the first quadrant we shall have the UES scenario, and then the blue points will be connected by a continuous curve; on the other hand, in the third quadrant we shall obtain the SSB scenario with disconnected points. For this instance, the qualitative behavior of the red lines, and the presence of a Mass Gap region will be maintained. This implies that the SSB-scenarios in the third quadrant are not stable, since all red lines will reach continuously the first quadrant (where $m^{2}>0$) at finite time, and thus the SSB-scenarios become UES-scenarios; in these cases we have that $lim_{t\rightarrow  +\infty} m^{2}(t,n)=+\infty$. The case of UES-scenarios in the first quadrant is more involved; consider by simplicity the case $n=2$, which is representative of all cases with $n\geq 2$ with 
$\mu_{0}>0$. The mass reads $m^2(n=2,t)=\frac{1}{-\mu t+\mu_{0}}$; thus, in the interval $[0,\frac{\mu_{0}}{\mu})$ the mass is increasing from a positive value at $t=0$; at $t=\frac{\mu_{0}}{\mu}$ a singularity appears, and in the interval  $(\frac{\mu_{0}}{\mu},+\infty)$, the mass is increasing from negative values, localized at the fourth quadrant. This abrupt change is not physical, since it can be eliminated by a reparametrization of $t$; under such a convenient reparametrization the original UES-scenarios are maintained stable.

The case with $\mu_{0}=0$ implies that the mass vanishes at $t=0$, and is well-defined only in the cases $n=0,-2,-4,...$; this case only presents the UES scenario, due to $m^{2}(t,n)\geq 0$. This case can be obtained from the Figure 1 by moving down the mass spectrum in the second quadrant in such a way the horizontal asymptote will be now the ``x-axis"; the red lines will  be essentially the same.

The change of the area corresponding to the manifold $M$ is given explicitly by $\partial_{t}{\cal A}=\omega\int_{M}(\frac{1}{2} m^{2}\phi^{2} + \frac{1}{16}\lambda\phi^{4})\Omega^{2l} d^{2}x$, according to Eq. (\ref{area}); thus in the UES scenario the area is increasing from certain initial value ${\cal A}_{0}$. Since in the case of SSB we have that $m^{2}<0$, the change in the area has no definite sign, and in general the area will have contractions and expansions, this could imply that $M$ may collapse to a point; explicit solutions for the Higgs  and metric flows  are required for determining the precise behavior of the area; this will be possible at vacuum, treated in the next section.

The evolution of the $\widehat{R}$-curvature and its traces is determined fully by the metric evolution (\ref{metric-flow}),
\begin{eqnarray}
     \partial_{t} \widehat{\Gamma}^{k}_{ij} \!\! & = & \!\!  - \omega [\delta^{k}_{(i} \partial_{j)}  - \frac{1}{2} \delta_{ij} \partial^{k}]V_{\Omega}, \quad
     g^{ij} \partial_{t} \widehat{\Gamma}^{k}_{ij}  = 0, \qquad \partial_{t} (\delta^{j}_{k} \widehat{\Gamma}^{k}_{ij})=-\omega\partial_{i}V_{\Omega}, \nonumber \\
     \partial_{t} \widehat{R} \!\! & = & \!\! \omega (\widehat{R} +  \Omega^{-2}{\Box})V_{\Omega},\quad
     \partial_{t}\widehat{R}_{ij}  =  \frac{\omega}{2} \delta_{ij}{\Box}V_{\Omega}, \quad
     \partial_{t} \widehat{R}^{i}{_{jkl}}  =  \omega \delta^{i}_{[k} \delta_{l]j}{\Box}V_{\Omega},          \label{evolution-6}
\end{eqnarray}
where $V_{\Omega}\equiv V\Omega^{2(l-1)}$, and $\Box\equiv\delta^{ij}\partial_{i}\partial_{j}$.
Similarly, the evolution of the $S$-curvature  is given by both the metric (\ref{metric-flow}) and the Higgs flows (\ref{Higgs-flow}),
\begin{eqnarray}
     \partial_{t} F_{ij} \!\! & = & \!\! 0, \quad
     \partial_{t} S  = \omega V_{\Omega}S  + \frac{\Omega^{-2}}{2} ({\Box} - \partial^{i}\phi\cdot\partial_{i}) (\frac{1}{2} \widehat{\Box}\phi - \partial_{\phi}V); \nonumber \\
     \partial_{t} R^{+}_{ij} \!\! & = & \!\! - \frac{1}{2} [\widehat{\nabla}_{(i} - \partial_{(i}\phi] \partial_{j)} (\frac{1}{2} \widehat{\Box}\phi - \partial_{\phi}V) +\frac{1}{2}\partial_{t} \Gamma^{k}_{ij}\cdot\partial_{k}\phi ;
      \label{evolution-7}
\end{eqnarray}
in this manner, the evolution of the $S$-curvature is determined by $\partial_{t}S^{i}{_{jkl}} =  2\delta^{i}_{[k}\partial_{t} R^{+}_{l]j}$. The stationary metric points in the case $\omega = 0$, induce stationary points for the $\widehat{R}$-curvature; however the geometry evolves according to (\ref{evolution-7}), along the unrestricted Higgs flow.

\section{Vacuum under evolution}

The evolution of the mass spectrum shown in the Figure 1, together with the diffusion-like equation (\ref{Higgs-flow}), and Eq. (\ref{metric-flow}) will determine the evolution of the configurations of the Higgs field  and of the metric at vacuum; this means that the geometries associated with vacuum will evolve in a nontrivial way. The vacua are defined by the vanishing of the derivative $\frac{\partial V(t)}{\partial \phi}=-\phi(t)[m^{2}(t)+\frac{\lambda}{4}\phi^2(t)]$, which depends on $t$.
In the case of UES the profile of a symmetric potential with a unique vacuum localized at $t_{0}$, $\phi_{vac}(t_{0},n)=0$, with $V_{vac}(t_{0},n)=0$, is not preserved along the flows for all $t$; Furthermore, the vanishing of the torsion
$T_{vac}(t_{0})=\nabla\phi_{vac}(t_{0})=0$, implies that the $S$-curvature vanishes, and consequently the unique vacuum corresponds to conformally flat manifolds with (usual) Riemannian geometry; such a vacuum is not stable under the Higgs-metric flows. Then, one requires of  explicit solutions of the coupled system for metric and Higgs flows
Eq. (\ref{metric-flow}), and Eq. (\ref{Higgs-flow}), in order to determine the evolution of such vacuum geometries, in particular if they will evolve to geometries with torsion; we will address this problem elsewhere.

In the case with SSB we have the profile of the potential with degenerated vacuum localized at $\phi_{vac}(t)=\pm2\sqrt{-\frac{m^{2}}{\lambda}}$ (Width) with $V_{vac}(t)=-\frac{m^{4}}{\lambda}$ (Height), it will have a nontrivial evolution determined by the evolution of the mass spectrum and of the parameter $\lambda$
described above. The metric flow (\ref{metric-flow}) reduces to $\partial_{t}\Omega^{2}_{vac}= \omega\frac{m^{4}}{\lambda}(\Omega^{2}_{vac})^{l}$, which is strictly positive; implying that the metric is expanding at vacuum for all $l$ .  The proportionality factor in this equation can be rewritten as $\omega\frac{m^{4}}{\lambda}=
\omega\frac{\lambda}{16}\phi_{vac}^{4}$, thus the solutions will respect the reflection symmetry.
 This equation for the conformal factor can be solved easily under the assumption of a stationary parameter  $\lambda$; however, positivity of a (real) conformal factor permits only the values $l\geq 0$, with restrictions on the mass spectrum; solutions for $l<0$ lead to complex expressions, and although the analytic extensions are of interest, we focus here only on real solutions. Therefore, the solutions are
\begin{eqnarray}
      \Omega^{2}_{vac}(x,t)\!\! & = & \!\! \left\{ \begin{array}{lll}\frac{\omega}{(n-3)\lambda \mu}(-m^{2})^{3-n}+\Theta, \qquad l=0, \quad n=4,6,8...,\\
\Theta\exp\{\frac{\omega}{(n-3)\lambda \mu}(-m^{2})^{3-n} \}, \quad l=1, \quad n=2,4,6,..,\\
    (l-1)^{\frac{1}{1-l}}[\frac{\omega}{\lambda \mu}(-m^{2})+\Theta]^{\frac{1}{1-l}} ,  \qquad l=3,5,7,..\quad n=2;
             \end{array} \right.
         \label{conformal-vacuum}
\end{eqnarray}
where $\Theta=\Theta(x)$, is a strict positive arbitrary function, and $\lambda$ is a (positive) constant. The asymptotic limit for $\omega>0$ of the conformal factor is given by
\begin{eqnarray}
     lim_{_{t\rightarrow  +\infty}} \Omega^{2}_{vac}(x,t)\!\! & = & \!\! \left\{ \begin{array}{lllll} +\infty , \qquad l=0, \quad n=4,6,8..,\\
     +\infty, \qquad l=1, \quad n=4,6,8..,\\
     \Theta, \qquad l=1, \qquad n=2,\\
         ((l-1) \Theta)^{\frac{1}{1-l}},  \qquad l=3,5,7,.., \quad n=2;   \\
                      \end{array} \right.
         \label{conformal-vacuum-asymp}
\end{eqnarray}
observe that the case $l=1$ has a splitting depending on the $n$-values; note also that the case $n=2$ of the mass spectrum is special, since if $\Theta$ is bounded, then is the only case with asymptotically bounded metric. Moreover, the area of the manifolds at vacuum is given directly by the expression ${\cal A}_{vac}(t)= \int_{M}\Omega^{2}_{vac}(x,t) d^{2}x$, and will inherit the asymptotic behavior from the conformal factor. The case $\omega=0$,  will correspond to a scenario with conformal factor and area preserved along the flows, determined essentially by the function $\Theta$.

Furthermore, in the Figure 2 we describe the evolution of the Higgs potential according to the evolution of its Width and Height; for simplicity we take $\lambda$ only as a stationary parameter. Under evolution, the potential becomes smoother, tending to eliminate the degeneration; however, this will take an infinite time.
\begin{figure}[H]
  \begin{center}
    \includegraphics[width=.4\textwidth]{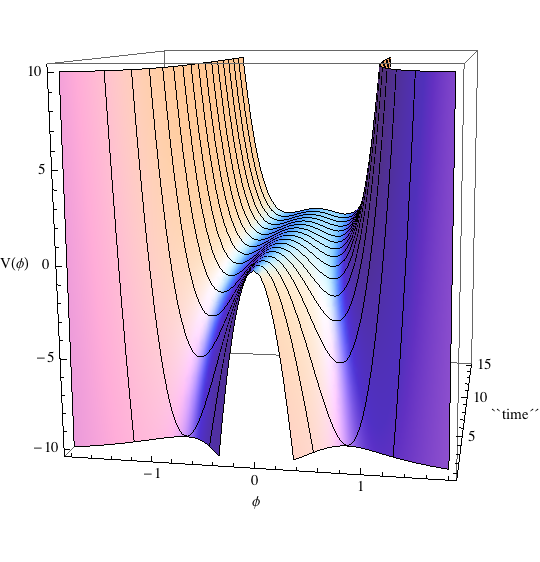}
  \caption{The Higgs potential under evolution: In the forefront we have the potential at $t=0$ with certain sharpness; this is soften under evolution, since both the Width and the Height of the barrier decrease  according to the behavior of the mass spectrum; this fact seems to favor the tunneling between the two minima. Since $lim_{t\rightarrow +\infty}[m(t,n)]=0$, then $lim_{t\rightarrow +\infty}[Width, Height]=0$, and the degeneration of vacuum is preserved at finite time; once the SSB has occurred, so is maintained along the flows.}
  \label{fig:ejemplo}
  \end{center}
\end{figure}
\newpage
As a result of the previous analysis, we can obtain that the quantum probability (per unit area) of tunneling between vacua is proportional to 
$$\Psi(t)\equiv\exp{[\int_{M} V_{vac}]}/{\cal A}_{vac}=\exp{[- \int_{M}\Omega^{2}_{vac}(\frac{m^{4}}{\lambda} )d^{2}x ]}/\int_{M}\Omega^{2}d^{2}x,
$$ 
which is even under reflection. From this expression we shall determine the qualitative aspects for possible transitions of manifolds, depending essentially on their boundedness.
We consider first the case with $\omega = 0$, which corresponds to manifolds of area determined by the integration over $M$ of the function $\Theta(x)$, preserved along the flows, and independent on the mass $m(t)$. At the beginning, $\Psi(t=0)$ is determined by the configuration of $m(t=0)$ (the blue points in the Figure 1), for each value admitted of $n$, which proves to be finite for manifolds with ${\cal A}$ bounded; thus, such a probability increases along the flow,  and reaches an asymptotic maximum as $lim_{t\rightarrow +\infty}\Psi(t,n,l)\sim \frac{1}{{\cal A}_{vac}}$. Therefore, in the case of two-manifolds at vacua with unbounded area, the probability of transition goes to zero for all $t$; in other words,  the manifolds with infinite area do not "climb the wall", and are concentrated around the minima, a result certainly expected. Note that this conclusion is valid in spite of the behavior of the Higgs potential in the Figure 2.

In the cases with $\omega>0$, the tunneling probability will be dominated by an increasing area along the flows;  the probability at $t=0$ will be given by the configuration of $m$ and ${\cal A}$ at $t=0$, and we assume that is finite for ${\cal A}$ finite. Now, in the first scenario with $l=0$, and $n=4,6,8,...,$ described in (\ref{conformal-vacuum-asymp}), the exponential is increasing for $n=4$, but bounded from above; thus  $lim_{t\rightarrow +\infty}\Psi(t,l=0, n=4)=0$, since the area is unbounded. In the same scenario, the cases with  $n=6,8,...,$, involve a decaying exponential, and the decay rate of $\Psi$ as $t\rightarrow +\infty$ is greater than the case $n=4$. The second scenario in (\ref{conformal-vacuum-asymp}) with $l=1$, and $n=4,6,8,...,$ is qualitatively similar. Qualitative differences of quantum tunneling emerge in the cases when the area is bounded along the flows and for $t\rightarrow +\infty$, which correspond to the last two cases in (\ref{conformal-vacuum-asymp}) described by the function $\Theta$.  Along the lines discussed above, we start with certain probability at $t=0$, it goes decreasing
but with lower limits given by $lim_{t\rightarrow +\infty}\Psi(t,l=1, n=2)\sim\frac{1}{\int_{M}\Theta}$, and $lim_{t\rightarrow +\infty}\Psi(t,l=3,5,7,..., n=2) \sim \frac{1}{\int_{M}((l-1) \Theta)^{\frac{1}{1-l}}}$, respectively. The last case deserves more discussion, since the probability has a maximum for certain combination of $l$ and the function $\Theta$. Consider first that  $\Theta$ is a constant (positive); thus, the  values of the probability against  $l$ lie on the following  curve (Figure 3), drawn for continuous values of $l$; this curve shows a maximum, although it is not localized precisely in an admissible value of $l$, since it depends also on $\Theta$; the corresponding value of $l$ is the closest to that calculated as a continuous variable, which maximizes the tunneling probability.

\begin{figure}[H]
  \begin{center}
    \includegraphics[width=.5\textwidth]{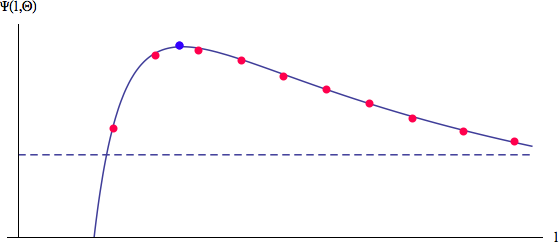}
  \caption{For $\Theta$ a positive constant, the maximum for the tunneling probability is localized at
the blue point in $\frac{e^{(l-1)^{2}}}{l-1}=\Theta e^{-1/\Theta}$, considering $l$ as a continuous variable.
 The admissible values $l=3,5,7,...,$ correspond to the red points; the red point closest to the blue point  gives the wished pair 
 $(l,\Psi)$. With the dashed line we show an additional limiting scenario with  $lim_{l\rightarrow +\infty}[\frac{1}{((l-1) \Theta)^{\frac{1}{1-l}}}]=\frac{1}{\Theta}$, which coincides with the  limit  for the case $(l=1, n=2)$ as ${t\rightarrow +\infty}$.}
  \label{fig:ejemplo}
  \end{center}
\end{figure}
If the function $\Theta$ depends on $x$, the condition for the maximum of the probability is given by the following integral constraint,
\begin{equation}
     \int_{M} \Theta^{\frac{1}{1-l}}[(l-1)^{2}+\frac{1}{\Theta}-\ln[(l-1)\Theta]] =0;
     \label{prob-max}
\end{equation}
once the integration is made, the constraint is reduced  to an algebraic equation for $l$ as a continuous variable; the value calculated allows to choice the closest $l$ among the admissible values.

In terms of the conformal factor the only component of the $\widehat{R}$-curvature is expressed as $\widehat{R}^{2}{_{112}} = -\frac{1}{2} \Box \ln \Omega^{2}$; at vacua it has the form
\begin{eqnarray}
    [\widehat{R}^{2}{_{112}}]_{vac} = - \frac{1}{2} \Box \left\{ \begin{array}{ll}
                                          \ln [\frac{\omega}{(n-3)\lambda\mu}(-m^{2})^{3-n} + \Theta], &  l=0, \quad n=4,6,8,\ldots ,\\
                                          \ln \Theta + \frac{\omega}{(n-3)\lambda\mu}(-m^{2})^{3-n}, & l=1, \quad n=2,4,6,\ldots, \\
                                          \frac{1}{1-l} \ln [\frac{\omega}{\lambda\mu} (-m^{2}) + \Theta], & l=3,5,7,\ldots \quad n=2;
                                          \end{array} \right.
                                          \label{R-vacuum}
\end{eqnarray}
which respects the reflection symmetry; this symmetry is maintained under evolution according to the Eq. (\ref{evolution-6}), since $( V_{\Omega})_{vac}$ is even under reflection. The asymptotic limit is given by
\begin{eqnarray}
     \lim_{t\rightarrow +\infty} [\widehat{R}^{2}{_{112}}] _{vac}= \left\{ \begin{array}{lll}
                                         \frac{1}{2} \Box \ln (\lambda \mu^{\frac{2}{n-1}}), & l=0, \quad n=4,6,8,\ldots \\
                                         \infty, & l=1, \quad n=4,6,8,\ldots \\
                                         -\frac{1}{2} \Box \ln \Theta , & l=1, \quad n=2, \\
                                         \frac{1}{2(l-1)} \Box \ln\Theta , & l=3,5,7,\ldots, \quad n=2,
                                         \end{array} \right.
                                         \label{R-asymp}
\end{eqnarray}
note that the first case of a divergent metric as $t\rightarrow +\infty$ in Eq. (\ref{conformal-vacuum-asymp}), is cured at the level of the $\widehat{R}$-curvature, due to the presence of the logarithmic function; in the second case the divergent behavior persists.

 Similarly to the UES case, the manifolds at the degenerate vacua have no torsion, $T_{vac}(t)=\nabla\phi_{vac}(t)=0$,  thus,
with a vanishing $S$-curvature along the flow, since
\begin{equation}
     \partial_{t} (R^{+}_{ij})_{vac} = \mp \frac{1}{4} (\widehat{\nabla}_{(i} \mp \partial_{(i} \phi_{vac}) \partial_{j)} \widehat{\Box} \phi_{vac} \pm \frac{1}{2} \partial_{t}( \widehat{\Gamma}^{k}_{ij})_{ vac} \cdot \partial_{k} \phi_{vac}=0.
     \label{S-vac-ev}
\end{equation}
Furthermore, considering that $\partial_{t}(\widehat{R}^{1}{_{112}})_{vac} = -\frac{\omega}{2} \Box (\frac{m^{4}}{\lambda} \Omega^{2l-2}_{vac})$, a negative gradient along the flows, then the geometries at vacua with positive curvature at $t=0$ tend to be flat, but bounded below according to Eq. (\ref{R-asymp}); by choosing appropriately the functions $\lambda$, $\mu$, and $\Theta$, such bound limits can be fixed to flat two-dimensional geometries. In the case of geometries with initial negative curvature, the flow tends to evolve to make it more negative, but  bounded from below. Thus, in general the flow tends to favor negative curvatures for vacua. 

\section{Concluding remarks}

The present scheme allows to distinguish dynamically between geometries with and without torsion; in the case with a SSB scenario,
the building of the perturbation theory around a ground state, corresponds to expand around conformally flat two-dimensional geometries without torsion (the usual ones), generating hence twisted geometries once the reflexion symmetry is broken spontaneously. These results have been obtained by using basically the parameters flows, ignoring explicit solutions for the diffusion-like equation for the Higgs field; explicit solutions are available in the literature, and we shall extend the results in future communications.

As commented previously, the explicit solutions for the coupled system of metric and Higgs flows will allow to determine the evolution of the potential in an UES scenario as initial condition; it is possible that there exist transitions between different scenarios, which will connect different phases of the theory by trajectories in the parameters space. 

String theory is formulated basically on the geometries associated to the vacua in the SSB scenario; it will be interesting
to re-formulate the theory using the twisted geometries that emerge by SSB.

\begin{center}
{\uno ACKNOWLEDGMENTS}
\end{center}
This work was supported by the Sistema Nacional de Investigadores (M\'{e}xico); the analysis of the differential equations and figures were made using Mathematica. RCF and AHA acknowledge support by a CONACYT Grant No. A1-S-38041, and JBM for financial support from CONACYT-Mexico under
the project No. CB-2017-283838.


\begin{thebibliography}{}
\setlength{\itemsep}{-.50em} \setlength{\itemsep}{-.50em}


\bibitem{Hamilton} R. S. Hamilton, J. Diff. Geo. \textbf{17}, 255-306 (1982). 

\bibitem{GFBHE}
J. Samuel and S. R. Chowdhury, 
Class. Quantum Grav. \textbf{24}, F47 (2007). 

\bibitem{EnergyERF}
J. Samuel and S. R. Chowdhury,  
Class. Quantum Grav. \textbf{25}, 035012 (2008). 

\bibitem{Dai} 
X. Dai and L. Ma, Commun. Math. Phys. \textbf{274}, 65 (2007). 

\bibitem{wiseman} M. Headrick and T. Wiseman, 
Class. Quantum Grav. \textbf{23}  6683 (2006). 

\bibitem{Viqar}
V. Husain and S. S. Seahra, Class. Quantum Grav. \textbf{25}, 222002 (2008). 

\bibitem{woolgar}
E. Woolgar, 
Can. J. Phys. \textbf{86}, 645 (2008). 

\bibitem{BahelowskyWoolgar}
T. Balehowsky, E. Woolgar, 
J. Math. Phys. \textbf{53}, 072501 (2012). 

\bibitem{figueras}
P. Figueras, J. Lucietti and T. Wiseman, 
Class. Quantum Grav. \textbf{28}, 215018 (2011).

\bibitem{kiritsis} 
E. Kiritsis, W. Li and F. Nitti,  Fortsch. Phys. \textbf{62}, 389 (2014). 

\bibitem{jackson} S. Jackson, R. Pourhasan and H. Verlinde, \textit{Geometric RG Flow,} arXiv:1312.6914 [hep-th], (2013). 

\bibitem{OSW1}
T. Oliynyk, V. Suneeta, and E. Woolgar,
Phys. Lett. B {\bf 610} 115 (2005).

\bibitem{OSW2}
T. Oliynyk, V. Suneeta, and E. Woolgar
Nucl. Phys. B {\bf 739} 441 (2006).

\bibitem{horava} P. Ho\v{r}ava, JHEP \textbf{0903}, 020 (2009). 

\bibitem{nishioka} T. Nishioka, Class. Quant. Grav. \textbf{26}, 242001 (2009). 

\bibitem{Chowdhury}
S. R. Chowdhury, Geometric flows and black hole entropy, \emph{ A Thesis submitted to the Jawaharlal Nehru University for the Degree of Doctor of Philosophy}, (2007).

\bibitem{perelman}
G. Perelman, {\it The entropy formula for the Ricci flow and its geometric applications}, arXiv:math.DG/0211159 (2002); {\it Ricci flow with surgery on three-manifolds}, arXiv:math.DG/0303109 (2003).

\bibitem{Oliynyk}
T. Oliynyk and E. Woolgar, 
\emph{Commun. Anal. Geom.}  \textbf{15}, 535 (2007).

\bibitem{RStability}
C. Guenther, J. Isenberg and D. Knopf, 
Commun. Anal. Geom \textbf{10}, 741 (2002).


\bibitem{Turck} 
D. M. DeTurck, J. Diff. Geom. \textbf{18}, 157 (1983). 

\bibitem{CFHAOS} 
R. Cartas-Fuentevilla, A. Herrera-Aguilar, J. A. Olvera-Santamaría, Eur. Phys. J. Plus \textbf{133}, 235 (2018). 

\bibitem{CHAHM}
R. Cartas-Fuentevilla, A. Herrera-Aguilar, and J. A. Herrera-Mendoza, Ann. Phys. \textbf{415}, 168093 (2020).

\bibitem{Taylor} 
M. Taylor, Class. Quantum Grav. \textbf{33} , 033001, (2016). 

\bibitem{HLS} 
S. A. Hartnoll, A. Lucas and S. Sachdev, \textit{Holographic quantum matter,} arXiv:1612.07324v3 [hep-th], (2018). 


\bibitem{cms-atlas}ATLAS Collaboration, Phys. Lett. B {\bf 716}, 1 (2012); CMS Collaboration, {\bf 716}, 30 (2012).

\bibitem{Tseytlin}
 A. A. Tseytlin, 
Phys. Rev. D {\bf 75}, 064024 (2007). 

\bibitem{OSW3}
T. Oliynyk, V. Suneeta, and E. Woolgar, 
Phys. Rev. D {\bf 76}, 045001 (2007).


\bibitem{cartas} R. Cartas-Fuentevilla, 
Int. Journal of Geom. Meth. Mod. Phys. {\bf 11}, 1450033 (2014).





\end{thebibliography}
\end{document}